\documentclass[conference]{IEEEtran}
\usepackage[utf8]{inputenc}           %
\usepackage{amsmath,amssymb,amsfonts}
\usepackage{graphicx}
\usepackage{textcomp}
\usepackage[pdftex,svgnames]{xcolor}
\usepackage{balance} %
\usepackage{tabularx}
\usepackage{paralist}
\usepackage{booktabs}
\usepackage[T1]{fontenc}
\usepackage[T1,lowtilde]{url}
\usepackage[bookmarks=false]{hyperref}
\usepackage[capitalize,noabbrev]{cleveref}

\usepackage[abbreviate=true, dateabbrev=true, alldates=year, isbn=false, doi=false, urldate=comp, url=false, maxbibnames=9, backref=false, backend=biber, style=numeric-comp, language=american, giveninits=true, sortcites=true, sorting=none]{biblatex}
\AtEveryBibitem{\clearlist{location}} %
\AtEveryBibitem{\clearfield{series}} %
\AtEveryBibitem{\clearlist{language}} %

\DeclareSourcemap{%
  \maps[datatype=bibtex]{
    \map{
      \step[fieldsource=booktitle, match={International}, replace={Int'l}]
      \step[fieldsource=booktitle, match=\regexp{[\{]{0,2}Conference[\}]{0,2}\s+on\s+}, replace=\regexp{Conf.\\\x{20}}]
      \step[fieldsource=booktitle, match=\regexp{[\{]{0,2}Symposium[\}]{0,2}\s+on\s+}, replace=\regexp{Symp.\\\x{20}}]
      \step[fieldsource=booktitle, match=\regexp{[\{]{0,2}Workshop[s]?[\}]{0,2}\s+on\s+}, replace=\regexp{Ws.\\\x{20}}]
      \step[fieldsource=booktitle, match={Software }, replace={Softw.\ }]
      \step[fieldsource=booktitle, match={Requirements }, replace={Req.\ }]
      \step[fieldsource=booktitle, match={Performance }, replace={Perf.\ }]
      \step[fieldsource=booktitle, match={Engineering}, replace={Eng.}]
      \step[fieldsource=booktitle, match={Advances\}\} and \{\{Innovations}, replace=\regexp{Adv.\\,\\\x{26}\\,Innov.}]
      \step[fieldsource=booktitle, match={Innovations}, replace={Innov.}]
      \step[fieldsource=booktitle, match={IFIP Conf.}, replace={IFIP Conf\}\}.}] 			
      \step[fieldsource=booktitle, match=\regexp{(.*)\s+\(.*\)$}, replace=\regexp{$1}]
      \step[fieldsource=journal, match=\regexp{(The\s+)?Journal\s+of\s+}, replace=\regexp{J.\\\x{20}}]
      \step[fieldsource=journal, match=\regexp{Communications\s+of\s+the\s+}, replace=\regexp{Comm.\\\x{20}}]
      \step[fieldsource=journal, match=\regexp{Engineering\s+Applications\s+of\s+(.*)}, replace=\regexp{Eng.\\\x{20}Applic.\\\x{20}$1}]
      \step[fieldsource=journal, match=\regexp{(ACM|IEEE)\s+Transactions\s+on\s+(.*)}, replace=\regexp{$1\x{20}Trans.\\\x{20}$2}]
      \step[fieldsource=journal, match={Information Security and Applications}, replace=\regexp{Inf.\\\x{20}Sec.\\\x{20}\\\x{26}\x{20}Applic.}]
      \step[fieldsource=journal, match={Applications}, replace={Applic.}]
      \step[fieldsource=journal, match={Programming}, replace={Progr.}]
      \step[fieldsource=journal, match=\regexp{(.*)\s+\(.*\)$}, replace=\regexp{$1}]
      \step[fieldsource=publisher, match={Academic }, replace={Acad.\ }]
      \step[typesource=techreport, typetarget=report]
    }
  }
}

\newcommand{\RemoveBibField}[2]{
  \DeclareSourcemap{\maps[datatype=bibtex]{\map{\step[fieldsource=entrykey, match={#1}, final]\step[fieldset=#2, null]
}}}}

\newcommand{\UseShortTitle}[1]{
  \DeclareSourcemap{\maps[datatype=bibtex]{\map{\step[fieldsource=entrykey, match={#1}, final]\step[fieldsource=title, match=\regexp{\A([^\:]+)\s*\:.+\z}, replace=$1]
}}}}

\RemoveBibField{tamura2013:practical}{editor}
\RemoveBibField{dias2020:visual}{editor}
\RemoveBibField{fleurey2009:domain}{editor}
\RemoveBibField{smith2009:methods}{editor}
\RemoveBibField{even-dar2006:dependable}{editor}
\RemoveBibField{linger1998:requirements}{pages}
\RemoveBibField{ghahremani2020:improving}{pages}
\RemoveBibField{porter2018:tess}{pages}
\RemoveBibField{sole2017:survey}{institution}
\RemoveBibField{sole2017:survey}{year}
\RemoveBibField{zhang2019:cnnfl}{pages}

\UseShortTitle{madni2018:modelbased}

\newcommand{\head}[1]{\par\noindent\textbf{#1:}\space}

\title{Adaptive Immunity for Software: \\ Towards Autonomous Self-healing Systems}

\author{%
\IEEEauthorblockN{%
Moeen Ali Naqvi \quad\quad%
Merve Astekin \quad\quad%
Sehrish Malik \quad\quad%
Leon Moonen\\[0.8ex]}
\IEEEauthorblockA{Simula Research Laboratory, Oslo, Norway}
\IEEEauthorblockA{Email:
\{moeen,merve,sehrish\}@simula.no, 
leon.moonen@computer.org}%
}

\makeatletter
\def\ps@IEEEtitlepagestyle{%
  \def\@oddfoot{\mycopyrightnotice}%
  \def\@evenfoot{}%
}
\def\mycopyrightnotice{%
  {\hspace*{3mm}\includegraphics[width=2cm]{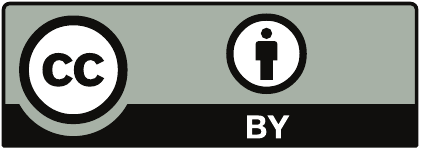}%
   \hspace*{2mm}\raisebox{2.5mm}{\parbox{\columnwidth}{\footnotesize This work is licensed under a Creative Commons \\ Attribution 4.0 International (CC BY 4.0) license.}}}
  \gdef\mycopyrightnotice{}%
}
\makeatother

\begin{document}

\maketitle

\noindent
\begin{abstract}
  Testing and code reviews are known techniques to improve the quality and robustness of software.
  Unfortunately, the complexity of modern software systems makes it impossible
  to anticipate all possible problems that can occur at runtime,
  which limits what issues can be found using testing and reviews.
  Thus, it is of interest to consider \emph{autonomous self-healing software systems},
  which can automatically detect, diagnose, and contain unanticipated problems at runtime.
  Most research in this area has adopted a model-driven approach,
  where actual behavior is checked against a model specifying the intended behavior,
  and a controller takes action when the system behaves outside of the specification.
  However, it is not easy to develop these specifications,
  nor to keep them up-to-date as the system evolves.
  We pose that, with the recent advances in machine learning,
  such models may be learned by observing the system.
  Moreover, we argue that artificial immune systems (AISs) are
  particularly well-suited for building self-healing systems,
  because of their anomaly detection and diagnosis capabilities.
  We present the state-of-the-art in self-healing systems and in AISs,
  surveying some of the research directions that have been considered up to now.
  To help advance the state-of-the-art,
  we develop a research agenda for building self-healing software systems using AISs,
  identifying required foundations, and promising research directions.
\end{abstract}

\begin{IEEEkeywords}
self-healing,
artificial immune systems,
anomaly detection,
runtime diagnosis,
fault containment,
dependability.
 \end{IEEEkeywords}

\section{Introduction}
\label{sec:introduction}

\noindent
Despite considerable investments in software verification and validation (V\&V),
we frequently read about software issues that cause severe damage:
airplanes have crashed, rockets have self-destructed, financial systems have stalled,
cars have been recalled, and patients have died from incorrect medication~\cite{comp.risks}.

Conventional V\&V techniques such as code reviews, static source code analysis, and (dynamic) software testing 
aim to ensure that software products satisfy their functional requirements and expected quality attributes.
However, one of the main challenges with these techniques is that they only target \emph{anticipated faults}:
it is inherent to their design that they can only check for \emph{known or expected} issues (or their generalizations)~\cite{kaner1999:testing}.
The increased complexity of modern software-intensive systems makes it difficult to anticipate
all possible problems that the system could encounter at runtime.

A recent study analyzed thirty prominent software failures since 2015 to examine how failures occur, 
and how they were detected, investigated and mitigated~\cite{sillito2020:failures}.
Main findings include that tests often fail to detect errors that are triggered by combinations of interactions,
and test environments often do not represent the production environment due to the scale and complexity of software in production.

Thus, we need complementary techniques that target the 
\emph{unanticipated faults} (anomalies) that remain after conventional V\&V.
A promising direction is that of \emph{autonomous self-healing software systems}.
Such systems monitor their own behavior, detect when an anomaly occurs, 
diagnose the root cause of the anomaly, and perform an intervention to contain it.

Although the terminology of self-healing systems draws on the healing of cells and tissue in the biological domain~\cite{ghosh2007:selfhealing}, 
the majority of research in this area does not use bio-inspired computing but a model-driven approach,
where \emph{actual} behavior is checked against a formal model specifying \emph{intended} behavior, 
and a controller takes action when the system behaves outside of the specification~\cite{bencomo2019:models}.
A significant drawback of this approach is that it is challenging to develop such models for large and complex software-intensive systems, 
and to keep them up-to-date as the system evolves~\cite{madni2018:modelbased}.

We argue that, with the recent advances in machine learning, 
such behavioral models need no longer be explicitly specified, 
but can be learned in a data-driven fashion, by observing the system in a suitably instrumented environment.
Moreover, we argue that \emph{artificial immune systems} (AISs), 
a machine learning paradigm inspired by the principles and processes of biological immune systems, 
are a particularly well-suited for building self-healing software systems,
because of their inherent focus on anomaly detection.
We take inspiration from earlier work that surveyed the application of AISs 
in the domain of manufacturing and process technology 
for the purpose of fault detection, diagnosis and recovery (FDDR)~\cite{bayar2015:fault}.
Their overall objectives are very similar to ours in building self-healing software systems, 
and the authors argue that also in that domain, 
the majority of existing methods are model-based diagnostic systems that are hard to implement due to their complexity.
They show how data-driven, quantitative methods such as AISs are well-suited for FDDR, 
which strengthens our belief that they will suit self-healing \emph{software} systems as well.

\head{Contributions}
We make the case for more research on data-driven self-healing software systems based on AISs.
To further stimulate this line of work, 
we survey the concepts and state-of-the-art in self-healing systems (\cref{sec:selfhealing}), 
as well as artificial immune systems (\cref{sec:artificialimmunesystems}).
Next, we develop a research agenda for self-healing systems using AISs (\cref{sec:rqs}), 
identifying required foundations and promising research directions,
and highlighting areas where the SANER community has specific expertise to further this domain.
Finally, we conclude in \cref{sec:conclusion}.

\section{Self-healing Systems}
\label{sec:selfhealing}

\begin{figure}
  \vspace*{1ex}%
  \centering
  \includegraphics[width=.5\columnwidth]{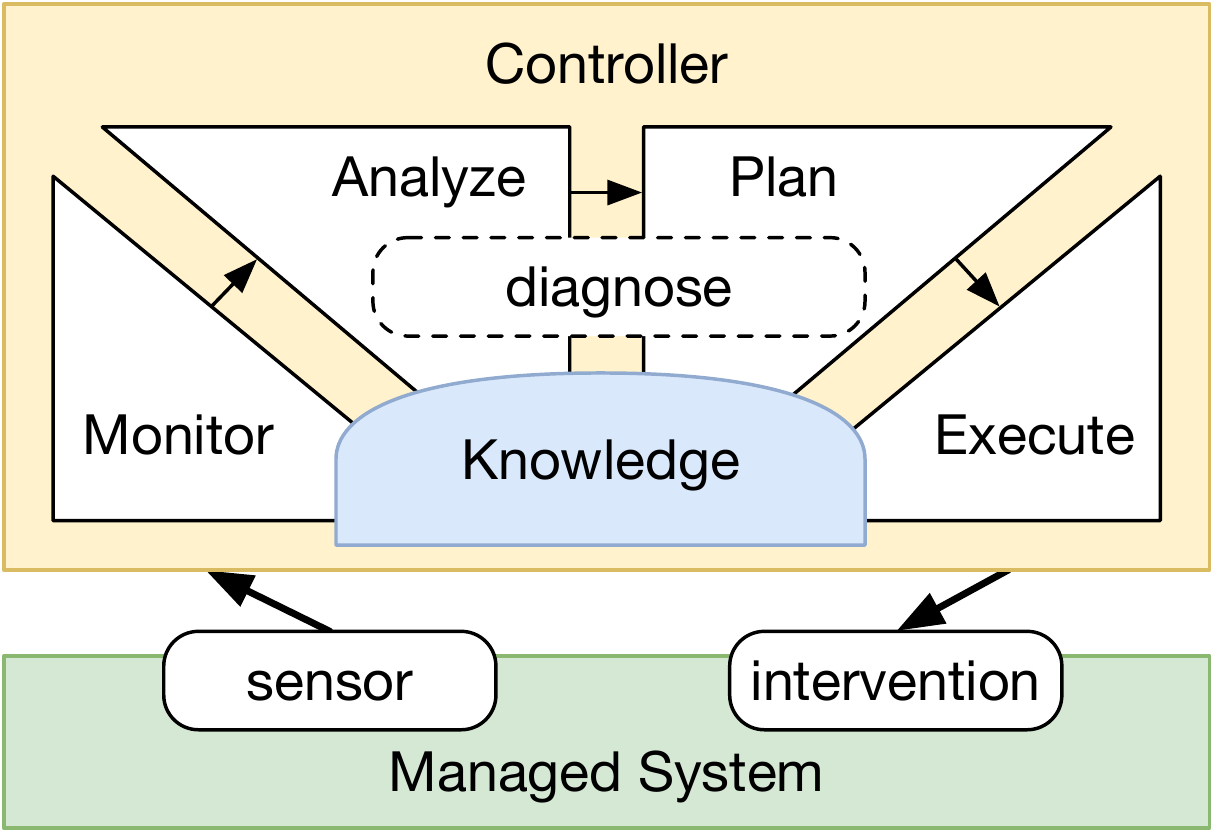}
  \vspace*{-1ex}%
  \caption{\label{fig:mape-k}MAPE-K feedback loop for self-healing systems (adapted from~\cite{kephart2003:vision}).}
  \vspace*{-2ex}%
\end{figure}

\head{Background}
Ganek et al. define self-healing systems as 
\emph{``systems [that] discover, diagnose, and react to disruptions. 
For a system to be self-healing, 
it must be able to recover from a failed component by first detecting and isolating the failed component, 
taking it off line, fixing or isolating the failed component, 
and reintroducing the fixed or replacement component into service without any apparent application disruption''}~\cite{ganek2003:dawning}.
The goal is improving system qualities such as robustness, 
availability, reliability and survivability~\cite{rajput2019:exploration}.
The concept evolved from various ideas introduced in the late 20th century, 
such as fault-tolerant systems~\cite{pierce1965:failuretolerant}, 
self-stabilizing systems~\cite{dijkstra1974:selfstabilizing}, 
survivable systems~\cite{linger1998:requirements}, 
self-adaptive systems~\cite{laddaga1999:creating}, 
and autonomic computing~\cite{ganek2003:dawning}. 
In addition, there are ideas with largely overlapping goals, 
such as intrusion-tolerant systems, recovery-oriented computing, and self-repair~\cite{rodhe2014:selfhealing}.
Self-healing systems are generally realized by extending a software system with a MAPE-K feedback loop, 
illustrated in Figure~\ref{fig:mape-k},
which \emph{Monitors} the system during execution, gathering data that exposes the actual behavior, 
\emph{Analyzes} this data to detect if the observed behavior violates desired behavior, 
and then \emph{Plans} and \emph{Executes} an intervention to contain the violation, 
based on \emph{Knowledge} about the system~\cite{kephart2003:vision}. 
The combination of Analysis and Planning is also referred to as \emph{diagnosis}.

\head{State-of-the-art} 
We focus on studies from the last decade, using keyword search, 
snowballing and reverse citation search.
Most of the studies use a \emph{model-based} approach,
which captures the knowledge about the desired behavior of the system, 
information needed to diagnose and plan which interventions are needed, 
and the architectures to enable interventions.
Bencomo et al. review the literature on model-based approaches at runtime~\cite{bencomo2019:models}. 
Their review covers a larger scope than just self-healing systems and includes 275 papers, 
where more than half of those target fault-tolerance and self-adaptation. 

There are two dominant approaches for analysis and planning: 
First, \emph{rule-based reasoning} approaches combine the two stages into a single diagnosis stage 
where interventions are executed when events occur that match specific rules or conditions.
The strengths include readability and the efficient processing of the rules, resulting in a scalable approach.
Farahani et al.\ present a self-healing architecture for industrial paint robots that uses rule-based reasoning~\cite{farahani2016:selfhealing}. 
The authors show an improvement in reliability and quality of the painting process. 
Drawbacks include that interventions are typically not optimal and rules have limited expressiveness~\cite{fleurey2009:domain}.

Second, \emph{utility-driven} approaches use optimization techniques to determine the optimal intervention strategy 
using a utility function that evaluates how valuable the result of each possible intervention is.
These approaches are characterized by optimal decisions, but the optimization limits scalability to large and complex configurations.
Ghahremani et al. present a hybrid that combines both approaches to achieve optimal decisions, 
while being scalable to large dynamic architectures~\cite{ghahremani2020:improving}.

\emph{Runtime verification} can be seen as a formal approach to 
developing model-based self-aware, and self-healing systems~\cite{leucker2009:brief}.
In runtime verification, the desired behavior is generally specified via logic formulae, 
often a form of linear temporal logic (LTL), 
and the monitor checks if the formulae hold for the events generated by the current execution.
Tamura et al.~\cite{tamura2013:practical} analyze the benefits, challenges, and concerns 
of using runtime verification to implement self-adaptive and self-healing systems.
Recent applications include a self-healing extension for Node-RED,
a visual programming solution for the Internet-of-things~\cite{dias2020:visual}, 
and a self-healing cloud computing infrastructure that uses a hybrid of logic formulae 
and `traditional' models for diagnosis and containment~\cite{alkasem2017:utility}.

\head{Evaluation}
The \emph{evaluation} of self-healing systems, i.e., 
the analysis whether proposed self-healing approaches work as intended,
is an area of particular significance for researchers. 
Graph grammars and graph transformations have been proposed as a way to formally model and verify the changes in self-healing and self-adaptive systems~\cite{bucchiarone2009:selfrepairing,eckardt2013:modeling}.
Vogel et al.~\cite{vogel2018:mrubis} propose an extensible simulator for 
model-based architectural self-healing systems called mRUBIS.
The system simulates an architectural runtime model of the system which can be manipulated by adaptation engines to execute (simulated) interventions.
It also supports injecting faults into the model which can be used for validation. 
Porter et al. propose a testbed for automatic generation of distributed software architectures 
and corresponding runtime applications~\cite{porter2018:tess}. 
The system collects variety of failure recovery and adaptation metrics 
which can be used for scientific analysis of the system. 
Ghahremani et al.~\cite{ghahremani2020:evaluation} reviewed the scientific literature 
to examine the state-of-the-art in evaluating self-healing systems. 
One of their main findings is that inputs used for evaluation are often not representative of the type of failures that the system is intended to handle.
They show that incorrect assumptions about the characteristics of failures result in large performance prediction errors, 
inadequate selection of self-healing approaches, and suboptimal parameter tuning.

\section{Artificial Immune Systems}
\label{sec:artificialimmunesystems}

\head{Background}
\emph{Artificial immune systems} (AISs) are a machine learning paradigm inspired by the principles and processes of biological immune systems. 
Like the biological immune system, AISs are adaptive, self-learning systems that can learn what is the normal situation, 
and use this knowledge to detect emergence of abnormal situations, i.e., anomalies, in real-time. 
The idea of using AISs to guard processes from undesired influences has been around since the 1980s~\cite{farmer1986:immune}. 
For most of that time, the available processing power and machine learning technology were not sufficient 
to adaptively learn the complex patterns involved in real systems, nor extrapolate them into more generally applicable ones. 
However, recent advances have increasingly enabled researchers 
to replicate the biological immune system's learning and memory capabilities for progressively complex tasks, 
and AISs have found applications in a wide range of domains~\cite{hart2008:application,dasgupta2014:artificial,bayar2015:fault}. 

Different features and analogies of the biological immune system have inspired a variety of AIS mechanisms. 
The main categories include: 
(1) immune response models, that lend themselves well to anomaly detection and diagnosis, 
(2) clonal selection algorithms, that are more geared towards (multi-objective) optimization problems, and
(3) immune network algorithms, which allow for more complex interactions that enable searching for solutions of problems such as classification, clustering and optimization~\cite{costasilva2015:survey}. 
In addition to these `pure' approaches, researchers have increasingly worked on hybrid approaches, 
that either enhance AISs with techniques such as fuzzy set theory, Bayesian inference, information theory, and kernel methods, 
or create ensembles of AISs with other machine learning paradigms, such as artificial neural networks, \emph{k}-Nearest Neighbors, and support vector machines~\cite{costasilva2015:survey}.

\begin{figure}
  \vspace*{1ex}%
  \centering
  \includegraphics[width=0.7\columnwidth]{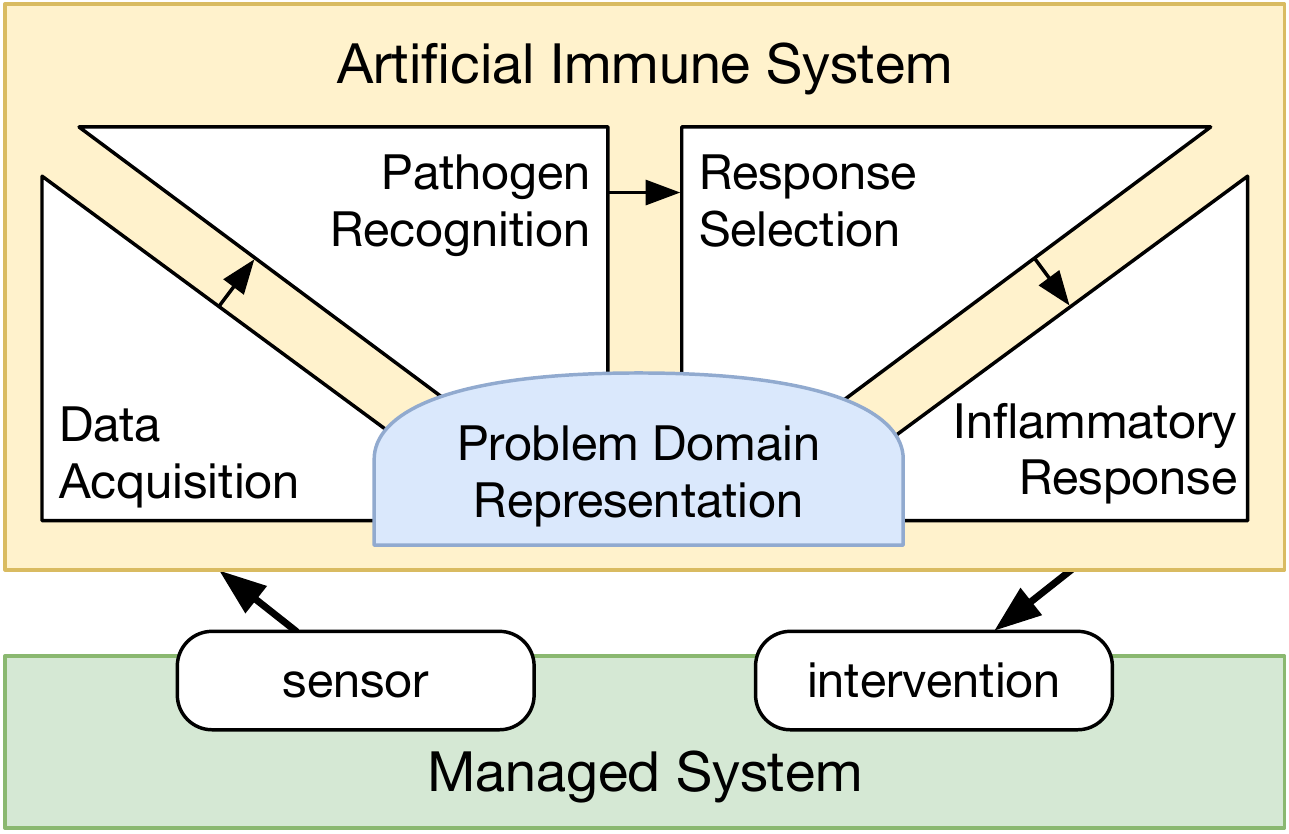}
  \vspace*{-1ex}%
  \caption{MAPE-K architecture for a self-healing artificial immune system.}
  \label{fig:architecture}
  \vspace*{-2ex}%
\end{figure}

\cref{fig:architecture} illustrates how the MAPE-K feedback loop can be instantiated for a self-healing AIS using an immune-response model.
The Knowledge is a representation of the problem domain that fits the AIS paradigm. 
A common representation is that of a collection of \emph{antibodies}. 
Each antibody is a tuple consisting of matching conditions and corresponding actions,
and antibodies have weighted connections to each other (known as their \emph{affinity}).
The Monitor acquires data from the sensors and encodes it for matching in the Analysis phase.
Analysis aims to recognize pathogens (anomalies) by computing an antibody ``concentration'' metric, 
based on the data acquired and the matching conditions and affinities of the antibodies.
Planning selects the antibody that has the highest concentration in the acquired data, 
and Execution responds with the actions corresponding to the selected antibody. 
This has two potential effects: it can trigger an intervention in the Managed System, 
and it can result in updates of the affinities between the selected and non-selected antibodies. 
This feedback loop serves to make the system adaptive and, for example, 
can be used to learn how to react more strongly to anomalies and less strongly to normal behavior.

\head{State-of-the-art}
We focus on studies that use AISs in a self-healing fashion.
Note that although familiar terms such as fault detection and fault diagnosis are used, 
most of these studies are from different domains than self-healing \emph{software} systems.

As mentioned in the introduction, Bayar et al. discuss the application of AISs for fault detection, 
diagnosis, and recovery (FDDR) in the domain of industrial manufacturing and process engineering~\cite{bayar2015:fault}. 
The authors show how data-driven, quantitative methods such as AISs are well-suited for FDDR, 
which strengthens our belief that they will suit self-healing \emph{software} systems as well.
They discuss the challenges and requirements of industrial FDDR, 
and classify the AIS mechanisms used in three categories of AISs with increasing complexity: 
(1) one-signal approaches based on self/non-self (SNS) discrimination. 
These are simple models (as sketched above) that select a single antibody as signal that a fault occurred, 
but as a result may suffer from false positives;
(2) two-signal approaches that combine SNS discrimination with the danger model of immune systems. 
This danger model establishes an overall danger signal based on all triggered antibodies, 
not just the one with the highest concentration.
In two-signal approaches, the first SNS signal is a necessary but not sufficient condition to raise an alarm, 
and a second danger signal is required as confirmation that a fault occurred, 
thereby reducing false positives; and
(3) immune network approaches, 
which are more complex configurations of the above, where fault information is propagated through a network representing the system. 
This not only allows for fault detection, 
but depending on the network configuration, also allows for fault localization, 
fault classification, or root cause analysis~\cite{bayar2015:fault}.

Immune networks have been used, for example, to address the problem of power system reconfiguration 
for effective service restoration in a smart self-healing power grid~\cite{oliveira2015:artificial}.
The AIS is used for locating and isolating the fault, as well as reconfiguring the network topology to reduce potential harm.

A considerable number of studies have focused on the applications of self-healing AISs in robotics~\cite{raza2015:immunoinspired}. 
A recent study proposes a robot immune system (RIS) as a basis for developing self-healing robots~\cite{akram2018:development}.
The RIS addresses both fault detection and fault recovery and its functionality was evaluated through a series of increasingly complex simulation scenarios 
using the Robot Operating System and Virtual Robot Experimentation Platform. 
Swarm robotics is an approach for the decentralized organization and coordination of a collection of relatively simple robots
with limited communication and interaction abilities.
Recent work proposed a new self-healing mechanism for robotic swarms that builds on the \emph{granuloma formulation} process in biological immune systems, 
which is the process of encapsulating pathogens in a protective layer to prevent the pathogen from spreading~\cite{timmis2016:immuneinspired}.
When applied in an AIS for swarm robotics, the granuloma formulation concept is used for fault containment by isolating defective robots.

The anomaly detection capabilities of AISs have been successfully applied in computer and network security~\cite{tan2016:artificial}:
Swimmer et al. use an AIS based on the \emph{danger model} for protecting a network from external threats such as computer viruses and worms, 
and argue that a two-signal based approach is essential in this context to achieve quick responses and reduce false positives~\cite{swimmer2007:using}.
Fernandes et al. survey applications of AISs to computer security, 
ranging from intrusion detection systems, network flooding/(D)DOS detection, fraud detection, spam detection, and phishing detection to malware detection~\cite{fernandes2017:applications}. 
The survey study concludes that AISs serve well as especially-tailored tools to address security issues, but their adoption in practice has been lacking.
A recent study by Aldhaheri et al. presents a comprehensive survey of AIS approaches to secure the Internet-of-Things (IoT)~\cite{aldhaheri2020:artificial}. 
The study covers various self-* aspects, such as self-learning, self-tolerance, self-adaptation, self-organizing and self-healing systems.  
They conclude that AIS approaches fit well in the context of dynamic environments such as IoT but scaling problems still exist. 
Some of the challenges they raise have to do with obtaining realistic datasets, and obtaining a realistic ground truth for evaluation.

\section{Research Directions}
\label{sec:rqs}

\newenvironment{questions}{\begin{compactitem}[\,~\textbullet]}{\end{compactitem}}

\noindent
When looking at the literature on self-healing systems,
we see that data-driven approaches such as AISs are underdeveloped 
in the context of self-healing \emph{software} systems.
While most AIS studies concentrate on other application domains,
the majority of research on self-healing software systems uses a model-driven approach.
However, these models are hard to develop for large and complex software-intensive systems, 
and are difficult to keep up-to-date as the system evolves~\cite{madni2018:modelbased}.

In this section, we focus on data-driven self-healing software systems based on AISs, 
and explore a number of promising research directions that can help to advance the state-of-the-art in AISs and self-healing software systems.

\head{Monitoring}
The appropriate data should be collected from managed systems~\cite{rabiser2017:comparison,candido2019:contemporary}. 
We see the following questions:
\begin{questions}   
    \item What observational data is needed for anomaly detection? 
    
    \item What are effective methods to collect this data?
    
    \item Can runtime observational data be augmented with software development data to increase detection accuracy?

\end{questions}
We believe the SANER community has specific expertise to answer the third question, 
for example, through its connections with earlier work on conceptual coupling.

\head{Anomaly Detection}
We have seen that AISs have successfully been used for anomaly detection in various domains, 
and their ability to adapt to changing environments may suit evolving software systems as well. 
The following questions arise when considering the use of AISs for self-healing software:
\begin{questions}
    \item To what extent can we successfully learn models of normal behavior purely from runtime observational data?   

	\item To what extent can we detect, and quantify, anomalies in the observational data to identify (potential) faults?
	
	\item Can we successfully recognize anomalies in a system that is being evolved and therefore changes behavior?

\end{questions}

\head{Fault Diagnosis}
\noindent
Fault diagnosis aims to identify the root causes of a detected anomaly, 
and investigate its impact on the system's components~\cite{sole2017:survey}.
This process becomes a challenging task for large and complex software-intensive systems,
due to their distributed, heterogeneous, and sometimes highly dynamic architecture~\cite{bennaceur2019:modelling}.
Yet, also for these systems we need to be able to accurately find the causes and potential effects of faults at runtime.
This raises the following questions:
\begin{questions}
    \item What are effective techniques to investigate and identify the root causes of a detected fault/anomaly? 
    \item What are successful strategies for root cause analysis in distributed, heterogeneous, and highly dynamic architectures, 
    	where diagnostic information may be scattered across a dynamically changing collection of components?
    \item What proactive actions can be taken to make the subject system easier to diagnose? At what cost?
\end{questions}
Starting points include diagnosis techniques such as \emph{spectrum-based fault localization} (SFL)~\cite{abreu2009:practical,wong2016:survey},
and more recent approaches based on convolutional neural networks~\cite{zhang2019:cnnfl}.

\head{Fault Recovery}
\noindent
A successful self-healing software system should mitigate faults and their observable side-effects to the extent that this is possible.
The recovery should include all affected components of the managed system to avoid cascading failures.
Moreover, the impact and risk of the recovery mechanisms should be assessed before applying it at runtime.
In this context, the following questions arise:
\begin{questions}
    \item What are the possible mechanisms to recover from a faulty state to a normal state? Which recovery techniques are effective for which level of the subject system?
    \item To what extent is it possible to recover from a faulty state? Is it sufficient to only address the original root cause of the failure, 
        or should the recovery process ``cascade'' to all affected components, similar to cascading failures?
    \item How is a recovery mechanism verified at runtime? To what extent can we learn patterns from failure records and take proactive actions for containment of faults?
\end{questions}
Starting points include earlier work on dependable computing and fault-tolerant architectures~\cite{even-dar2006:dependable,knight2012:fundamentals,smith2009:methods}

\head{Evaluation}
\noindent
An important research area in the domain of self-healing software systems is evaluation of the self-healing technology itself.
Proper validation of self-healing technology requires realistic datasets and, ideally, systems that can be shared as test subjects, 
so that approaches can be bench-marked and compared.
This raises the following questions:
\begin{questions}
    \item How, and at what scale, can realistic faults scenarios be modeled and executed to enable effective and repeatable evaluation and validation of self-healing technology?
    \item To what extent do we need real world observational data and failure traces, or can we simulate or synthesize these?
    \item What are effective methods for generating the data needed to evaluate and validate self-healing technology? 
\end{questions}
The SEAMS community has recently started collecting so-called \emph{examplars} that could form a starting point for this line of work~\cite{2020:exemplars:web}.

\section{Concluding Remarks}
\label{sec:conclusion}

\noindent
This paper argues for more research on \emph{data-driven} self-healing software systems, 
and identifies \emph{artificial immune systems} as a particularly well-suited paradigm 
because of its inherent anomaly detection and diagnosis capabilities.
We explore the state-of-the art in self-healing systems as well as artificial immune systems, 
and identify a series of promising research directions, 
highlighting initial research questions and connections to expertise in the SANER community.

\smallskip

\head{Acknowledgment}
This work is supported by the Research Council of Norway through the cureIT project (\#300461). 
Sehrish Malik is supported through the ERCIM `Alain Bensoussan’ Fellowship Programme.

\balance
\printbibliography 

\end{document}